\documentclass[twocolumn,amsmath,amssymb]{revtex4}
\usepackage{graphicx}
\usepackage{dcolumn}
\usepackage{bm}
\begin{document}

\title{Effective tuning of electron charge and spin distribution in a
dot-ring nanostructure at the ZnO interface}
\author{Tapash Chakraborty $^1$,\footnote{Tapash.Chakraborty@umanitoba.ca}
Aram Manaselyan$^2$, and Manuk Barseghyan$^2$} \affiliation{$^1$
Department of Physics and Astronomy, University of Manitoba,
Winnipeg, Canada R3T 2N2} \affiliation{$^2$ Department of Solid
State Physics, Yerevan State University, Yerevan, Armenia e-mails:
amanasel@ysu.am, mbarsegh@ysu.am}
\date{\today}
\begin{abstract}
Electronic states and the Aharonov-Bohm effect in ZnO quantum
dot-ring nanostructures containing few interacting electrons reveal
several unique features. We have shown here that in contrast to the
dot-rings made of conventional semiconductors, such as InAs or GaAs,
the dot-rings in ZnO heterojunctions demonstrate several unique
characteristics due to the unusual properties of quantum dots and
rings in ZnO. In particular the energy spectra of the ZnO dot-ring
and the Aharnov-Bohm oscillations are strongly dependant on the
electron number in the dot or in the ring. Therefore even small
changes of the confinement potential, sizes of the dot-ring or the
magnetic field can drastically change the energy spectra and the
behavior of Aharonov-Bohm oscillations in the system. Due to this
interesting phenomena it is possible to effectively control with
high accuracy the electron charge and spin distribution inside the
dot-ring structure. This controlling can be achieved either by
changing the magnetic field or the confinement potentials.
\end{abstract}

\maketitle

Self-assembled semiconductor quantum nanostructures, such as quantum
dots (QDs) and quantum rings (QRs) have been investigated
extensively given their potential as building blocks for novel
optoelectronic devices, e.g. nanoemitters, efficient solar cells,
terahertz detectors and for quantum information technologies
\cite{Chakraborty1,Michler,Li,Fomin}. The potential of these
nanostrcutures is based on their remarkable similarity to atomic
systems. Furthermore, what makes these nanostructures so attractive
is the ability to tune their optoelectronic properties by carefully
designing their size, composition, strain and shape. In this
context, QRs with their doubly-connected structure attracts special
attention. Its unique topological structure provides a rich variety
of fascinating physical phenomena in this system. Observation of the
Aharonov-Bohm (AB) oscillations \cite{Aharonov} and the persistent
current \cite{Buttiker} in small semiconductor QRs, and recent
experimental realization of QRs with only a few electrons
\cite{Lorke,haug} have made QRs an attractive topic of experimental
and theoretical studies for various quantum effects in these
quasi-one-dimensional systems \cite{Chakraborty}.

Recently we have witnessed an increasing demand for the realization
of complex quantum confined systems, such as laterally coupled QDs,
QRs and also quantum dot-ring (QDR) complexes
\cite{Meier,Kiravittaya,Somaschini}, for both practical applications
and fundamental studies, including geometrical quantum phase
\cite{Capozza}, spin-spin interactions \cite{Saiga} and quantum
state couplings \cite{daSilva}. Due to the unique topology of the
QDR structures and their potential applications, the theoretical
investigation of QDR's properties has received much attention in
recent years. A few-electron system confined in a QDR in the presence
of an external magnetic field shows that the distribution of electrons
between the dot and the ring is influenced by the relative strength
of the dot-ring confinement and the magnetic field which induces
transitions of electrons between the two parts of the system \cite{Szafran}.
These transitions are accompanied by changes in the periodicity of
the AB oscillations. It has been recently shown \cite{Zipper,Zipper1,Kurpas1}
that many measurable properties of a QDRs, such as spin relaxation or optical
absorption, can be significantly changed by modifying the confinement
potentials, which demonstrates the high controllability and
flexibility of these systems. The transport properties of QDR nanostructures
are also known to be drastically modified due to the unique geometry
\cite{Kurpas2}. Theoretical studies of the dc current through a QDR in the
Coulomb blockade regime have revealed that it can efficiently work as a
single-electron transistor or a current rectifier. In our recent works the
electronic and optical properties of QDRs with hydrogenic donor
impurity were investigated in external electric and magnetic fields
\cite{Manuk1,Manuk2,Manuk3}

In the past few years, very exciting developments have taken place
with the creation of high-mobility 2DEG in heterostructures
involving insulating complex oxides \cite{ZnO}. Unlike in
traditional semiconductors, electrons in  these systems are strongly
correlated \cite{mannhart}. These should then exhibit effects
ranging from strong electron correlations, magnetism, interface
superconductivity, tunable metal-insulator transitions, among
others, and of course, the exciting possibility of all-oxide
electronic devices. Many surprising results were found in the
fractional quantum Hall states \cite{fqhe} discovered in the
MgZnO/ZnO heterojunction \cite{falson,luo}, in particular, in tilted
magnetic fields \cite{tilted,falson,luo_2}. Preparation of various
nanostructures, such as nanorings, nanobelts, etc. have been
reported in ZnO \cite{zno_nano}. In our earlier reports, electronic,
optical and magnetic properties of ZnO QDs and QRs were investigated
\cite{ZnOQD,ZnOQR}. We have shown that for both systems the
electron-electron interaction effects are much stronger then in
traditional semiconductor quantum systems, such as InAs or GaAs. On
the other hand the AB effect in a ZnO QR strongly depends on the
electron number \cite{ZnOQR}. In fact, for two electrons in the ZnO
ring, the AB oscillations become aperiodic, while for three
electrons the AB oscillations completely disappear. Therefore,
unlike in conventional quantum ring topology, here the AB effect
(and the resulting persistent current) can be controlled by varying
the electron number. Taking into account these surprising and
interesting results, here we report on our studies of the electronic
states and the AB effect in a ZnO QDR containing few electrons. We
have shown that in such systems it is possible to control with high
accuracy the electron charge and spin inside the dot or the ring at the single-%
electron level.

\begin{figure}
\includegraphics[width=4cm]{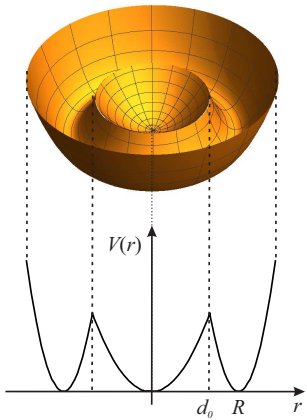}
\caption{\label{fig:EdepB1} The schematic picture of QDR confinement
potential.}
\end{figure}

Our present study involves a two dimensional quantum dot-ring
structure with cylindrical symmetry, based on the 2DEG at the ZnO
interface, containing few interacting electrons, in a magnetic field
that is applied in the growth direction. The Hamiltonian of our
system is
\begin{equation}
{\cal H}=\sum_i^{N^{}_e}{\cal H}_\mathrm{SP}^i+\frac12\sum_{i\neq
j}^{N^{}_e}V^{}_{ij}, \label{Ham2D}
\end{equation}
where $N^{}_e$ is the number of electrons in the QDR,
$V^{}_{ij}=e^2/\epsilon\left|\mathbf{r}^{}_i
-\mathbf{r}^{}_j\right|$ is the Coulomb interaction term, with
dielectric constant of the material $\epsilon$, and ${\cal
H}^{}_\mathrm{SP}$ is the single-particle Hamiltonian in the
presence of an external perpendicular magnetic field.
\begin{equation}\label{Hsp}
{\cal H}^{}_\mathrm{SP}=\frac{1}{2m}\left(\textbf{p}-\frac
ec\textbf{A}\right)^2+V^{}_{\mathrm{conf}}(r)+\frac12
g\mu^{}_BB\sigma^{}_z,
\end{equation}
where $\textbf{A}=B/2(-y,x,0)$ is the vector potential, and $m$ is
the electron effective mass. The last term of (\ref{Hsp}) is the
Zeeman splitting. We choose the confinement potential of the QDR
consisting of double parabolas \cite{Manuk3,Szafran}.
$V^{}_{\mathrm{conf}}(r)=\min\left[\frac{1}{2}m\omega_{\mathrm{d}}^2r^2,
\frac12m\omega_{\mathrm{r}}^2(r-R)^2\right], $ where $\omega^{}_{\rm
d}$ and $\omega^{}_{\rm r}$ are the parameters describing the
strength of the confinement potential and also the sizes of the dot
and the ring respectively. The radius of the ring $R$ is defined as
the sum of oscillator lengths for the dot and ring related wells and
the barrier thickness $d$ between dot and ring according to
$R=\sqrt{2\hbar /m\omega^{}_d}+\sqrt{2\hbar /m\omega^{}_r}+d$. In
Fig.~1 the QDR confinement potential is presented schematically.

The eigenfunctions and the eigenenergies of the single-electron Hamiltonian
(\ref{Hsp}) can be obtained with the help of the exact diagonalization technique,
with the basis of wave functions of the cylindrical QD with larger radius.
We have used the exact diagonalization scheme also to calculate the energy
spectra and wave functions of few-electron QDR. This method is very accurate
and widely used by many authors \cite{Chakraborty1}. In order to
evaluate the energy spectrum of the many-electron system, we need to
digonalize the matrix of the Hamiltonian (\ref{Ham2D}) in a basis of
the Slater determinants constructed from the single-electron wave
functions \cite{ZnOQR}. In our model we have used 132 single-%
electron basis states. As a result we got 8646 two-electron states
and 374660 three-electron basis states which is adequate for
determining the first few energy eigenvalues for each value of the
total angular momentum of electrons with high accuracy.

In order to determine the average electron numbers in the dot or in
the ring we have also studied the electron densities for few-%
electron states in the QDR $\rho(\textbf{r})=\int
d\textbf{r}^{}_2d\textbf{r}^{}_3\ldots
|\Psi^{}_i(\textbf{r},\textbf{r}^{}_2,\textbf{r}^{}_3,\ldots)|^2, $
where $\Psi^{}_i(\textbf{r},\textbf{r}^{}_2,\textbf{r}^{}_3,\ldots)$
is the wave function of the few-electron state $i$. For the average
electron number in the dot region of QDR we get
$
N^{}_{\mathrm{dot}}=\int_0^{d^{}_0}\int_0^{2\pi}\rho(\textbf{r})rdrd\varphi,
$
where
$d^{}_0=\omega^{}_{\mathrm{r}}R/(\omega^{}_{\mathrm{d}}+\omega^{}_{\mathrm{r}})$
is the radius of the border between the dot and the ring.

\begin{figure*}
\includegraphics[width=10cm]{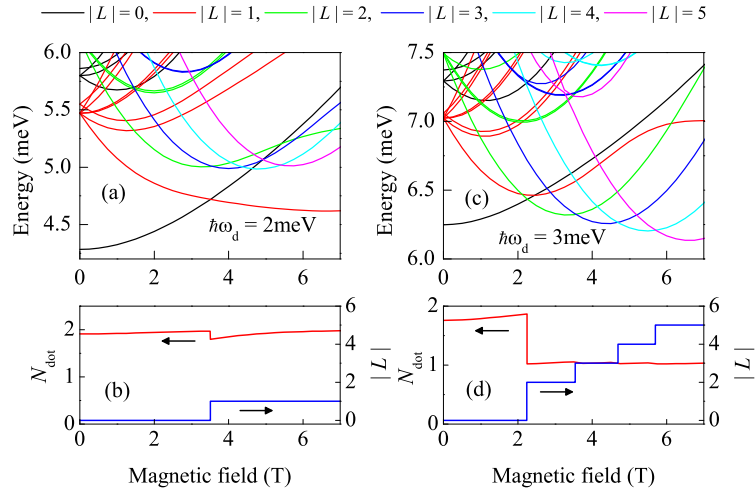}
\caption{Magnetic field dependencies of two-electron energy spectra,
the corresponding average electron number in the dot (left scales of
lower row) and the ground state angular momenta (right scales of lower
row) for $\hbar\omega^{}_d=2$ meV (a),(b) and for
$\hbar\omega^{}_d=3$ meV(c),(d). All results are for
$\hbar\omega^{}_r=8$ meV and $d=10$ nm.}
\end{figure*}

Our investigations were carried out for the ZnO QDR with parameters
$m=0.24m^{}_0$, $g=4.3$, $\epsilon=8.5$ \cite{Handbook}. In
Fig.~2(a) The magnetic field dependence of the first few energy
levels are presented for two-electron QDR with
$\hbar\omega^{}_{\mathrm{d}}=2$ meV, $\hbar\omega^{}_{\mathrm{r}}=8$
meV, $d=10$ nm for various values of the total angular momenta $L$.
In Fig.~2(b) corresponding ground state average electron number in
the dot (left scale, red line) and the ground state angular momentum
(right scale, blue line) are presented. From these figures it is
clear that for all values of the magnetic field, both electrons are
localized in the dot region and the ground state behaves like in a
two-electron single QD \cite{ZnOQD}. For weak values of the magnetic
field the ground state is a singlet state with angular momentum
$L=0$ and total spin $S=0$. With the increase of the magnetic field,
at $B\approx3.5$T, a singlet-triplet transition of the ground state
is observed to the state $L=-1$, $S=-1$. Similar results are
presented also in Fig.~2(c) and (d) but for
$\hbar\omega^{}_{\mathrm{d}}=3$ meV. In this case again for weak
values of the magnetic field both electrons are located in the dot
region, but starting from $B\approx2.2$T one of the electrons moves
to the ring region (Fig.~2(d) red line) and the ground state changes
to the triplet state $L=-2$, $S=-1$. Starting from $2.2$T the
behavior of the energy spectra in Fig.~2(c) changes drastically and
the regular periodic AB oscillations of the ground state can be
observed with periodic change of the total angular momentum by
$\Delta L=1$. This result is qualitatively similar to the case of
one-electron ZnO QR, observed in our previous work \cite{ZnOQR}, but
in contrast to that now all AB oscillations occur between the triplet states.

\begin{figure*}
\includegraphics[width=10cm]{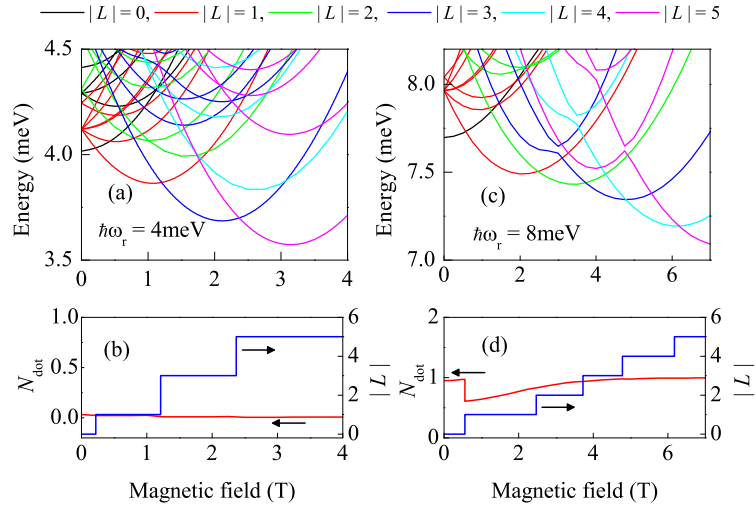}
\caption{Magnetic field dependencies of two-electron energy spectra,
corresponding average electron number in the dot (left scale of
lower row) and ground state angular momenta (right scales of lower
row) for $\hbar\omega^{}_r=4$ meV (a),(b) and for
$\hbar\omega^{}_r=8$ meV(c),(d). All results are for
$\hbar\omega^{}_d=4$ meV and $d=10$ nm.}
\end{figure*}

Figure~3 is the same as Fig.~2 but for a fixed value of
$\hbar\omega^{}_{\mathrm{d}}=4$ meV and for two values of the ring
confinement parameter $\hbar\omega^{}_{\mathrm{r}}=4$ meV (Fig.~3(a)
and (b)) and $\hbar\omega^{}_{\mathrm{r}}=8$ meV (Fig.~3(c) and
(d)). For $\hbar\omega^{}_{\mathrm{r}}=4$ meV both electrons are
localized in the ring (Fig.~3(b)) and the energy spectra is similar
to the one of the two-electron ZnO QR previously observed in
\cite{ZnOQR}. In this case irregular AB oscillations are observed
which is typical for ZnO QRs. Due to the combined effect of the
strong Zeeman splitting and the strong Coulomb interaction in ZnO,
the singlet-triplet crossings disappear from the ground state. For
small magnetic fields the ground state is a singlet with $L=0$. With
an increase of the magnetic field the ground state changes to a
triplet with $L=-1$ and $S=-1$. With further increase of the
magnetic field all the observed crossings of the ground state
correspond to triplet-triplet transitions between the states with
odd number of total angular momentum ($|L|=1,3,5...$). With the
increase of $\hbar\omega^{}_{\mathrm{r}}$ one of the electrons moves
to the dot region (Fig.~3(d)) while the other remains in the ring.
In this case the magnetic field almost does not change the average
electron number in the dot, and therefore we observe almost regular
AB oscillations similar to ZnO QR with one electron (Fig.~3(c)).

\begin{figure*}
\includegraphics[width=15cm]{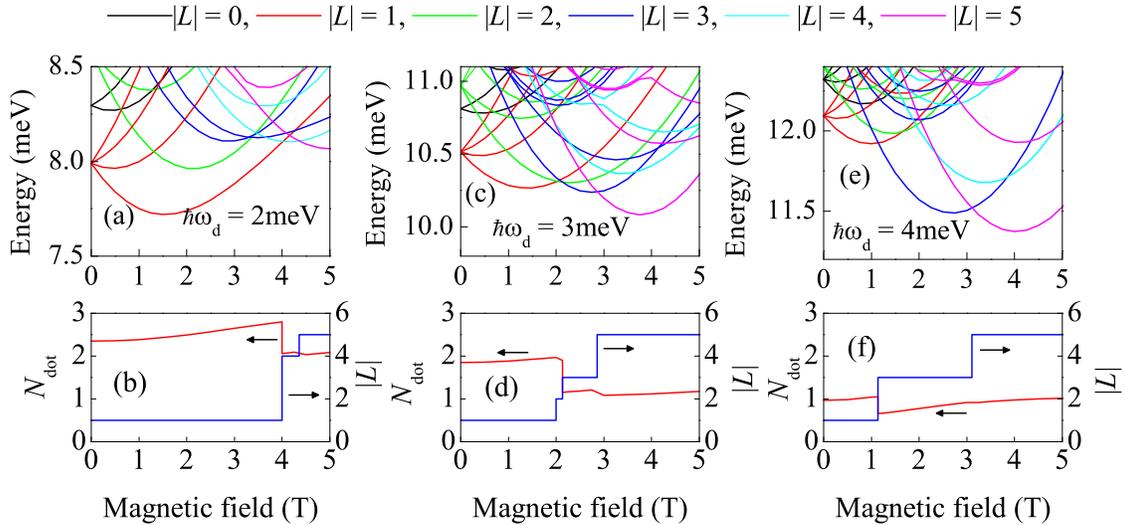}
\caption{Magnetic field dependencies of three-electron energy
spectra, corresponding average electron number in the dot (left
scales of lower row) and ground state angular momenta (right scales
of lower row) for $\hbar\omega^{}_d=2$meV (a),(b), for
$\hbar\omega^{}_d=3$meV(c),(d) and for
$\hbar\omega^{}_d=4$meV(e),(f). All results are for
$\hbar\omega^{}_r=8$meV and $d=10$nm.}\end{figure*}

The energy spectra and average electron numbers in the dot region are
presented in Fig.~4 against the magnetic field for a QDR containing three
electrons for fixed value of $\hbar\omega^{}_{\mathrm{r}}=8$ meV and for
various values of $\hbar\omega^{}_{\mathrm{d}}$. For $\hbar\omega^{}_{\mathrm{d}}
=2$ meV (Fig.~4(a) and (b)) and for weak magnetic fields all three electrons
are mostly located in the dot region, the ground state is $L=-1$, $S=-1/2$
and AB oscillations are not observed. But at $B\approx4$T one of the electrons
moves to the ring region and the ground state changes to $L=-4$, $S=-3/2$. Starting from
$B\approx4$T the usual AB effect appears.

For $\hbar\omega^{}_{\mathrm{d}}=3$ meV (Fig.~4(c) and (d)) without
the magnetic field, only two electrons are located in the dot and
one in the ring region. With an increase of the magnetic field we
can observe a charge switching between the dot and the ring of the
QDR. At $B\approx2.1$T the electron number in the dot changes to 1
and the irregular AB effect is observed with odd angular momenta
$|L|=3,5...$.

Finally, for $\hbar\omega^{}_{\mathrm{d}}=4$ meV (Fig.~4(e)
and (f)) and for all observed range of the magnetic field, only one electron
is located in the dot region and the other two stays in the ring. Therefore,
in Fig.~4(e) the irregular AB oscillations are observed for the ground
state, similar to the case of two-electron ZnO QR, as we found earlier
\cite{ZnOQR}.

To summarize: we have studied the electronic states and the
Aharonov-Bohm effect in ZnO quantum dot-ring nanostructures
containing few electrons. We have shown that in contrast to QDRs of
conventional semiconductors, such as InAs or GaAs, QDRs in ZnO
heterojunctions demonstrate several unique characteristics. In
particular the energy spectra of the ZnO QDR and the Aharnov-Bohm
oscillations are strongly dependant on the electron number in the
dot or in the ring. Therefore even small changes of the confinement
potential, sizes of the dot-ring or magnetic field can drastically
change the energy spectra and the behavior of Aharonov-Bohm
oscillations in the system. Due to this interesting phenomena it is
possible to effectively control with high accuracy the electron
charge and spin distribution inside the dot-ring structure. These
unique properties will certainly have important implications for
possible applications in spintronic devices and quantum information
technologies.

The work was supported by the Canada Research Chairs Program of the
Government of Canada, and Armenian State Committee of Science
(Project no. 15T-1C331).

\end{document}